\documentclass[twocolumn,prl,aps,twocolumn]{revtex4}
\usepackage[latin9]{inputenc}
\usepackage{amsmath}
\usepackage{amssymb}
\usepackage{graphicx}
\usepackage{makecell}
\usepackage{braket}
\usepackage{indentfirst}
\usepackage{mathrsfs}

\usepackage[unicode=true,pdfusetitle,
 bookmarks=false,
 breaklinks=false,pdfborder={0 0 1},backref=false,colorlinks=false]
 {hyperref}
\hypersetup{
 bookmarksnumbered=false,bookmarksopen=false}

\makeatletter
\@ifundefined{textcolor}{}
{%
 \definecolor{BLACK}{gray}{0}
 \definecolor{WHITE}{gray}{1}
 \definecolor{RED}{rgb}{1,0,0}
 \definecolor{GREEN}{rgb}{0,1,0}
 \definecolor{BLUE}{rgb}{0,0,1}
 \definecolor{CYAN}{cmyk}{1,0,0,0}
 \definecolor{MAGENTA}{cmyk}{0,1,0,0}
 \definecolor{YELLOW}{cmyk}{0,0,1,0}
}

\usepackage{color}

\newcommand{\ehbar}{\hbar_{\mathrm{eff}}}

\@ifundefined{textcolor}{}{%
 \definecolor{BLACK}{gray}{0}
 \definecolor{WHITE}{gray}{1}
 \definecolor{RED}{rgb}{1,0,0}
 \definecolor{GREEN}{rgb}{0,1,0}
 \definecolor{BLUE}{rgb}{0,0,1}
 \definecolor{CYAN}{cmyk}{1,0,0,0}
 \definecolor{MAGENTA}{cmyk}{0,1,0,0}
 \definecolor{YELLOW}{cmyk}{0,0,1,0}
}

\usepackage{txfonts}
\begin{document}

\title{Dynamical stability in a non-Hermitian kicked rotor model}

\author{Wen-Lei Zhao}
\email{wlzhao@jxust.edu.cn}
\affiliation{School of Science, Jiangxi University of Science and Technology, Ganzhou 341000, China}
\author{Huiqian Zhang}
\affiliation{School of Science, Jiangxi University of Science and Technology, Ganzhou 341000, China}

\begin{abstract}
We investigate the quantum irreversibility and quantum diffusion in a non-Hermitian kicked rotor model for which the kicking strength is complex. Our results show that the exponential decay of Loschmidt echo gradually disappears with increasing the strength of the imaginary part of non-Hermitian driven potential, demonstrating the suppress of the exponential instability by non-Hermiticity. The quantum diffusion exhibits the dynamical localization in momentum space, namely, the mean square of momentum increases to saturation with time evolution, which decreases with the increase of the strength of the imaginary part of the kicking. This clearly reveals the enhancement of dynamical localization by non-Hermiticity. We find, both analytically and numerically, that the quantum state are mainly populated on a very few quasieigenstates with significantly large value of the imaginary part of quasienergies. Interestingly, the average value of the inverse participation ratio of quasieigenstates decreases with the increase of the strength of the imaginary part of the kicking potential, which implies that the feature of quasieigenstates determines the stability of wavepacket's dynamics and the dynamical localization of energy diffusion.
\end{abstract}
\date{\today}

\maketitle

\section{introduction}

Quantum irreversibility and energy diffusion are two aspects of the fundamental problem of quantum chaos~\cite{Haake10}. Quantum mapping models provide ideal platforms for investigating quantum chaoticity from different prospectives, such as the eigenenergy level spacing and the wavepacket dynamics. A paradigm model of quantum mapping systems is the quantum kicked rotor (QKR), which has been widely employed in the study of the fundamental problems, for instance quantum-classical transition~\cite{WGWang12}, quantum irreversibility~\cite{WGWang05}, ergodicity~\cite{Cao22}, and prethermalization~\cite{Martinez22}. The landmark study by Peres shown that for classically chaotic systems, the small perturbation on the Hamiltonian leads to the exponential divergence of the fidelity, i.e., Loschmidt echo, between two nearby quantum states~\cite{Peres}, which is a solid evidence of exponential instability of quantum chaos. For the quantum diffusion, a seminal phenomenon in the QKR model is the dynamical localization (DL) in momentum space~\cite{Casati79}, which was later theoretically proven to be an analogy of Anderson localization in disordered lattices~\cite{Fishman82}.
The finding of DL spurs extensive investigations, both theoretically and experimentally, on the exotic diffusion phenomena in variants of QKR model~\cite{Santhanam22}, on the exponential instability in the presence of perturbation~\cite{WGWang04}, and on the dynamical phase transition in momentum space lattice~\cite{Hainaut18}.

In recent years, much interest has been focused on non-Hermitian systems, where novel phenomena~\cite{Zhou2,Gong1,Zhang1,Li2}, such as anomalous topology~\cite{HuH}, quantum entanglement~\cite{Gopalakrishnan}, and phase transitions~\cite{PiresDP} have been found. This triggered extensive attentions in diverse fields of physics, for instance electric circuits~\cite{Bergholtz,Zou}, atomic-optical setting with gain and loss~\cite{YongmeiXue}, quantum metrology~\cite{Budich,McDonald}, as well as open quantum systems~\cite{Dai,XiaoL}. It is found that in a $\cal{PT}$-symmetric extension of the QKR model the spontaneous $\cal{PT}$ symmetry breaking emerges with a scaling law depending on both the strength of the imaginary part of the kicking potential and the dimension of the system~\cite{West10}. In the regime of the breaking phase of $\cal{PT}$ symmetry, the $\cal{PT}$-symmetric kicked rotor exhibits the quantized acceleration in momentum space~\cite{Zhao1}, and the quantized response of out-of-time-ordered correlators with respect to the variation of the kicking strength~\cite{Zhao22}, which enriches our understanding on the fundamental problems of quantum transport and information scrambling in non-Hermitian chaotic systems. More interestingly, the mean-field treatment of many-body interaction even leads to the superexponential diffusion of energy in the QKR with non-Hermitian driven potential~\cite{Zhao20}.

In this context, we investigate the quantum irreversibility and quantum diffusion in a non-Hermitian kicked rotor (NQKR) model. It is known that in Hermitian case the Loschmidt echo decays exponentially with time with a rate proportional to the Lyapunov exponent of classical chaos. We find that the non-Hermitian kicking suppresses the exponential decay of Loschmidt echo, which even remains at unity for sufficiently strong non-Hermitian kicking strength, signalling the disappearance of irreversibility in the NQKR model. The quantum diffusion in momentum space displays the phenomenon of DL, representing by the emergence of the exponentially-localized wavepacket and the saturation of mean energy, which is dramatically reduced by increasing the strength of non-Hermitian driven potential. According to the Floquet theory, we predict that a quantum state will finally evolve to a quasieigenstate with large most value of the imaginary part of quasienergy, which is verified by our numerical results of the fidelity between the time-evolved quantum state and quasieigenstates. The feature of exponential localization of quasieigenstates determines the quantum stability of the NQKR model. We numerically make statistical measurement on the exponential localization of the quasieigenstates, and find that the inverse participation ratio (IPR) increases with increasing the real part of the kicking strength and decreases with the increase of its imaginary part, which demonstrates the enhancement of DL by non-Hermiticity.

The paper is organized as follows. In Sec.~\ref{DynStab}, we discrible the NQKR model and show the dynamical stability. In Sec.~\ref{EDLocali}, we present the enhancement of DL by non-Hermitian driven potential. In Sec.~\ref{SEC-Mecha}, we reveal the mechanism of dynamical stability and DL in the NQKR model. Conclusion and discussion are presented in Sec.~\ref{SEC-Conclusion}.

\section{Dynamical stability induced by non-Hermitian driven potential}\label{DynStab}

The Hamiltonian of a NQKR model in dimensionless units reads
\begin{equation}\label{NonHamil}	\textup{H}=\frac{p^2}{2}+V_{K}(\theta)\sum\limits_{n=-\infty}^{+\infty}\delta(t-t_n)\;,
\end{equation}
where the complex kicking potential is in the form $V_{K}(\theta)=\left(K+i\lambda\right)\cos(\theta)$ with $K$ and $\lambda$ indicating the strength of the real and imaginary parts, respectively. Here, $p=-i\ehbar\partial/\partial\theta$ is angular momentum operator, $\theta$ denotes the angular coordinate, satisfying the commutation relation $\left[\theta, p\right] = i\ehbar$ with $\ehbar$ the effective Planck constant. In the basis of the angular momentum operator, an arbitrary state can be expanded as $|\psi\rangle=\sum_{n}\psi_{n}|\varphi_{n}\rangle$, where $|\varphi_{n}\rangle$ is the eigenstate of $p$, $p|\varphi_{n}\rangle=p_n|\varphi_{n}\rangle$, with eigenvalue $p_n=n\ehbar$. One period time evolution of a quantum state is governed by $|\psi(t_{n+1})\rangle =U|\psi(t_n)\rangle$, where the Floquet operator is in the form
\begin{equation}\label{FlqOptr} U=\exp\left(-\frac{i}{\ehbar}\frac{p^{2}}{2}\right)\exp\left[-\frac{i}{\ehbar}V_{K}(\theta)\right]\;.
\end{equation}
In numerical simulations, we choose the ground state as the initial state, i.e., $\langle \theta|\psi(t_0)\rangle=1/\sqrt{2\pi}$.

A common measure of the instability of quantum dynamics is the Loschmidt echo
\begin{equation}\label{Loshecho}
\mathcal{L}(t)=|\langle\psi(t_0)|\exp(i\textrm{H}t/\hbar)
\exp(-i\textrm{H}_{\epsilon}t/\hbar)|\psi(t_0)\rangle|^2\;,
\end{equation}
where $\textrm{H}_{\epsilon}=\textrm{H} +\epsilon V$ represents a perturbed Hamiltonian on the original one $\textrm{H}$~\cite{Peres,Haug,Shres,Probst}. It is known that for chaotic systems, the Loschmidt echo exponentially decays with time, for which the decay rate is proportional to the Lyapunov exponent of the classical limits. Accordingly, the dynamics of $\mathcal{L}$ reflects a kind of quantum-classical correspondence of chaotic systems. For the Hermitian QKR, the classical dynamics is governed by the kick strength $K$, which is fully chaotic for $K>1$ with the Lyapunov exponent $\gamma \approx \ln(K/2)$~\cite{Chirikov79}.
\begin{figure}[t]
\begin{center}
\includegraphics[width=7.5cm]{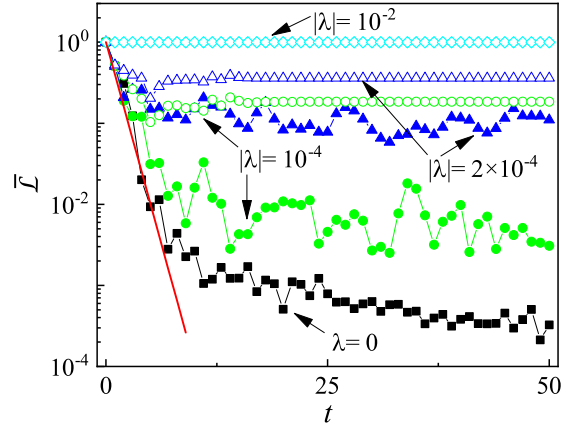}
\caption{ (a) The $\bar{\mathcal{L}}$ versus time with $\epsilon=10^{-3}$ for $\lambda =0$ (squares), $\pm 10^{-4}$ (circles), $\pm 2\times10^{-4}$ (triangles), and $\pm 10^{-2}$ (diamonds). Solid (empty) symbols represent positive (negative) $\lambda$. Note that the $\mathcal{L}$ for $\lambda =\pm 10^{-2}$ almost completely overlaps with each other. Red line indicates the exponential decay $\bar{\mathcal{L}}\propto e^{-\gamma t}$ with the Lyapunov exponent $\gamma=\ln(K/2)$. The parameters are $K=5$ and $\ehbar=3\times 10^{-5}$.}\label{Expinstal}
\end{center}
\end{figure}

We numerically investigate the time evolution of $\mathcal{L}$ of the NQKR model for $K>1$ with focus on the chaotic dynamics of Hermitian case. In numerical simulations we choose the Gaussian wavepackets as the initial state, i.e., $\psi{(\theta_c,t_0)}=(\sigma/\pi)^{1/4} \exp [-\sigma (\theta-\theta_c)^{2}/2]$ with $\sigma=10$. In addition, we numerically calculate the average of $\mathcal{L}$ for different $\psi{(\theta_c,t_0)}$, i.e., $\bar{\cal{L}}(t)=\sum_{j=1}^N \mathcal{L}_j(t)/N$ with $\mathcal{L}_j(t)=|\langle\psi(\theta_c^j,t_0)|\exp(i\textrm{H}t/\hbar)
\exp(-i\textrm{H}_{\epsilon}t/\hbar)|\psi(\theta_c^j,t_0)\rangle|^2$ and $\theta_c^j=2\pi j/N$, so as to reduce the dependence of $\mathcal{L}$ on the initial states. Figure~\ref{Expinstal}(a) shows that for Hermitian case, i.e., $\lambda=0$, the $\bar{\mathcal{L}}$ exponentially decays, i.e., $\bar{\mathcal{L}}\sim e^{-\gamma t}$ from unity to saturation, with $\gamma$ being the Lyapunov exponent. Interestingly, for small $\lambda$ (e.g., $\lambda=|10^4|$), the $\bar{\mathcal{L}}$ follows that of Hermitian case for very short time duration, after which it saturates with the saturation value being apparently larger than that of Hermitian case. For medium $\lambda$, e.g., $\lambda=|2\times10^4|$, the time interval for the exponential decay of $\mathcal{L}$ is very small, and the saturation value of $\mathcal{L}$ with $\lambda=-2\times 10^4$ is clearly larger than that of $\lambda=2\times10^4$ demonstrating a kind of asymmetry in the NQKR system. Interestingly, for large $\lambda$ (e.g., $\lambda=|10^2|$), the $\mathcal{L}$ almost remains at unity with time evolution, which clearly demonstrates the disappearance of irreversibility of the quantum dynamics induced by non-Hermiticity.
\begin{figure}[htbp]
\begin{center}
\includegraphics[width=8.5cm]{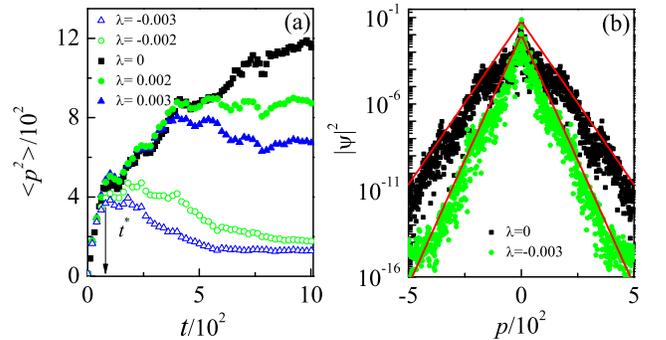}
\caption{(a) The  $\langle p^{2}\rangle$ versus time with $\lambda=-0.003$ (empty triangles), -0.002 (empty circles), 0 (squares), 0.002 (solid circles), and 0.003 (solid triangles). Arrow marks the threshold time $t^*$. (b) Momentum distributions at the time $t_n=1000$ for $\lambda=0$ (squares) and -0.003 (circles). Solid lines indicates the exponential function $|\psi(p)|^2\propto e^{-|p|/\xi}$ with $\xi\approx23$ and 15 for $\lambda=0$ and -0.003, respectively. Other parameters are $K=5$ and $\ehbar=0.25$.}\label{Ener}
\end{center}
\end{figure}

\section{Enhancement of dynamical localization by non-Hermitian driven potential}\label{EDLocali}

It is known that the DL which is an analogy of Anderson localization~\cite{Anders1} emerges in Hermitian QKR due to quantum coherence~\cite{Qd-Satpathi}. In order to investigate the features of DL in NQKR, we take numerical experiments on simulating the time evolution of  the mean square of momentum $\langle p^2\rangle=\sum_{n}p_n^2|\psi_n|^2/\cal{N}$ for different $\lambda$, where $\mathcal{N}=\sum_{n}|\psi_n|^2$ is the norm of a quantum state. This kind of definition of $\langle p^2\rangle$ eliminates the contribution of the norm to expectation value, which will exponentially increase for large $\lambda$. Figure~\ref{Ener}(a) shows that for $\lambda=0$ there is a short time interval for energy diffusion, after which the quantum mean energy gradually approaches to saturation, signalling the onset of DL. For nonzero $\lambda$ (e.g. $\lambda=0.002$), the $\langle p^2\rangle$ follows that of $\lambda=0$ for finite time $t^*$, beyond which it saturates. Interestingly, both the critical time $t^*$ and the saturation value of $\langle p^2\rangle$ decrease with the increase of $|\lambda|$. Therefore, the increase of the strength of non-Hermiticity dramatically enhances the DL of energy diffusion. Detailed observation shows that both the $t^*$ and the saturation value of mean energy for $-\lambda$ are smaller than that of $\lambda$, which demonstrates the asymmetry of DL in this system. We further investigate the momentum distributions for different $\lambda$. It is known that for $\lambda=0$, the DL is accompanied by the appearance of the exponentially localized wavepacket in momentum space $|\psi(p)|^2\sim e^{-|p|/\xi}$ with the localization length $\xi$ being almost unchanted with time [see Fig.~\ref{Ener}(b)]. For nonzero $\lambda$ (e.g., $\lambda=-0.003$), the quantum state exhibits the exponentially-localized shape, for which the $\xi$ is smaller than that of $\lambda=0$. This presents a clear evidence of the enhancement of DL by non-Hermiticity from the probability density distribution point of view.

\begin{figure}[htbp]
\begin{center}
\includegraphics[width=7.5cm]{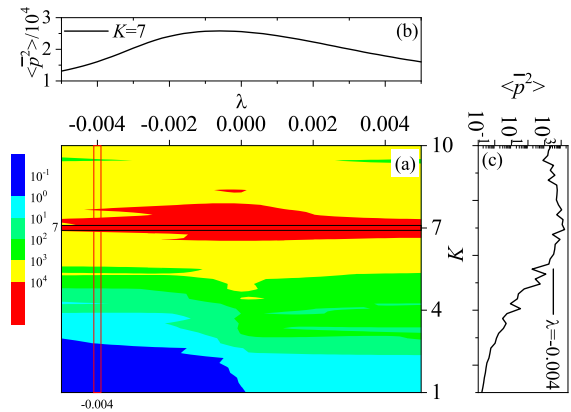}
\caption{(a) The time-averaged value of mean energy $\langle \bar{p}^{2}\rangle$ in the parameter space $(K,\lambda)$ with $\ehbar=0.25$. (b) The $\langle \bar{p}^{2}\rangle$ versus $\lambda$ with $K=7$. (c) The $\langle \bar{p}^{2}\rangle$ versus $K$ with $\lambda=0.004$.}\label{avEner}
\end{center}
\end{figure}

The saturation value of mean energy can be well quantified by the time-averaged value $\langle \bar{p}^2\rangle=\sum_{n=1}^{N}\langle p^2(t_n)\rangle/N$ for $N\gg 1$. We numerically investigate the $\langle \bar{p}^2\rangle$ for a wide regime of $K$ and $\lambda$. In numerical simulations, we find that a thousand of kicks $N=1000$ is large enough to assure the well approximation of the saturation level of $\langle p^2\rangle$ by $\langle \bar{p}^2\rangle$, so long as $K$ is not too large and $\ehbar$ is not too small. The phase diagram of $\langle \bar{p}^2\rangle$ in Fig.~\ref{avEner}(a) displays clearly different regime of diffusion behavior in the parameter space. For the a specific $K$, [e.g., $K=7$ in Fig.~\ref{avEner}(b)], the value of $\langle \bar{p}^2\rangle$ decreases with the increase of $|\lambda|$, and is asymmetric with respective to the change $\lambda\rightarrow -\lambda$, which coincides the enhancement of DL by $\lambda$ in Fig.~\ref{Ener}. For a fixed $\lambda$ [e.g. $\lambda=0.004$ in Fig.~\ref{avEner}(c)], the value of $\langle \bar{p}^2\rangle$ increases with the increase of $K$, demonstrating the assistance of quantum diffusion by kicking strength. Our results may be helpful for guiding the experimental investigations in the field of atom-optics with non-Hermiticity~\cite{YongmeiXue}.

\section{Mechanism of the enhancement of dynamical localization by non-Hermiticity}\label{SEC-Mecha}

Floquet theory predicts the eigenequation $U|\varphi_{\varepsilon}\rangle = e^{-i\varepsilon}|\varphi_{\varepsilon}\rangle$, where $|\varphi_{\varepsilon}\rangle$ is the quasieigenstate and $\varepsilon$ indicates the corresponding quasienergy~\cite{Shirley,Sambe}. In the basis of $|\varphi_{\varepsilon}\rangle$, an initial state can be expanded as $|\psi(t_0)\rangle = \sum_{\varepsilon}C_{\varepsilon}|\varphi_{\varepsilon}\rangle$. According to the Floquet theory, one can straightforwardly get the expression $|\psi(t_n)\rangle = \sum_{\varepsilon}C_{\varepsilon}e^{-i\varepsilon t_n}|\varphi_{\varepsilon}\rangle$. Note that in our system, the quasienergy is complex $\varepsilon=\varepsilon_r + i\varepsilon_i$ when $\lambda$ is sufficiently large, so we have the expansion $|\psi(t_n)\rangle = \sum_{\varepsilon}C_{\varepsilon}e^{-i\varepsilon_r t_n}e^{\varepsilon_i t_n}|\varphi_{\varepsilon}\rangle$~\cite{Longhi17}. It is clearly that the components with positive $\varepsilon_i$ exponentially grow and that with negative $\varepsilon_i$ exponentially decay.
This implies that the time-evolved state gradually approaches to the $|\varphi_{\varepsilon}\rangle$ with large most $\varepsilon_i$, and
dynamics of the quantum state $\psi(t_n)$ is governed by these quasieigenstates.

\begin{figure}[htbp]
\begin{center}
\includegraphics[width=8.5cm]{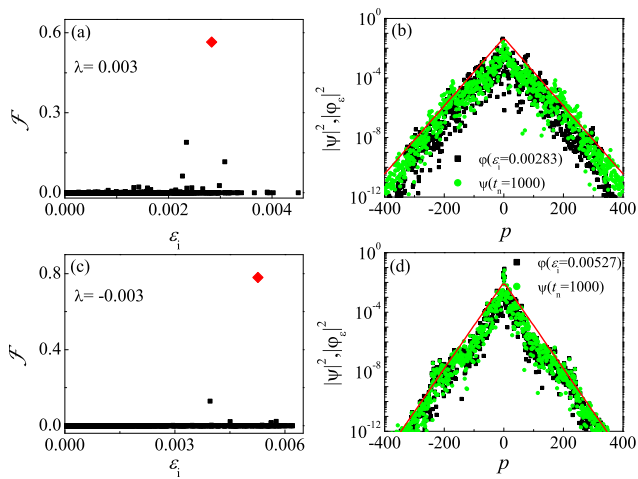}
\caption{Left panels: Dependence of $\mathcal{F}$ at the time $t_n=1000$ on the imaginary part of the quasienergy $\varepsilon_{i}$ with $\lambda= 0.003$ (a) and -0.003 (c). Right panels: Comparison of the probability density distributions between the state $|\psi(t_n = 1000)\rangle$ (circles) and the quasieigenstate $|\varphi_{\varepsilon}\rangle$ of the maximum value of $\mathcal{F}$ (red diamonds) with $\lambda= 0.003$ (b) and -0.003. (d) Red lines indicate the exponentially-localized shape $|\psi(p)|^2\propto e^{-|p|/\xi}$ with $\xi\approx19$ (b) and 15 (d). Other parameters are the same as in Fig.~\ref{Ener}(a).}\label{fiderity}
\end{center}
\end{figure}

To confirm this conjecture, we numerically investigate the fidelity between the time-evolved quantum state and the quasieigenstates, i.e, $\mathcal{F}(t_n)=|\langle\psi(t_n)|\varphi_{\varepsilon}\rangle|^2$ for the time in the regime of DL. Figure~\ref{fiderity}(a) shows that for $\lambda=0.003$, the $\mathcal{F}$ at the time $t_n=1000$ is nonzero for $\varepsilon_i >0$ and has a maximum value at $\varepsilon_i = 0.00283$. As a further step, we compare the probability density distribution between quantum state $|\psi(t_n=1000)\rangle$ and quasieigenstate $|\varphi_{\varepsilon}\rangle$ with $\varepsilon_i = 0.00283$ in Fig.~\ref{fiderity}(b). One can find that the two state almost completely overlap with each other, both of which are exponentially localized in momentum space, i.e., $|\psi(p)|^2\propto e^{-|p|/\xi}$ with $\xi\approx 19$. It is reasonable to believed that the exponential localization of quasieigenstates leads to the DL of the dynamics of quantum diffusion. For $\lambda=-0.003$, the maximum $\mathcal{F}(t_n=1000)$ corresponds to $\varepsilon_i=0.00527$ [see Fig.~\ref{fiderity}(c)]. Both the quantum state $|\psi(t_n=1000)\rangle$ and the quasieigenstate $\varphi_{\varepsilon}(\varepsilon_i=0.00527)$ display the same exponentially-localized shape in momentum space with the localization length $\xi\approx 15$ [see Fig.~\ref{fiderity}(d)]. It is apparent that the exponentially-localized feature of quasieigenstates determines the DL of the spreading of quantum states in momentum space.

\begin{figure}[htbp]
\begin{center}
\includegraphics[width=8.5cm]{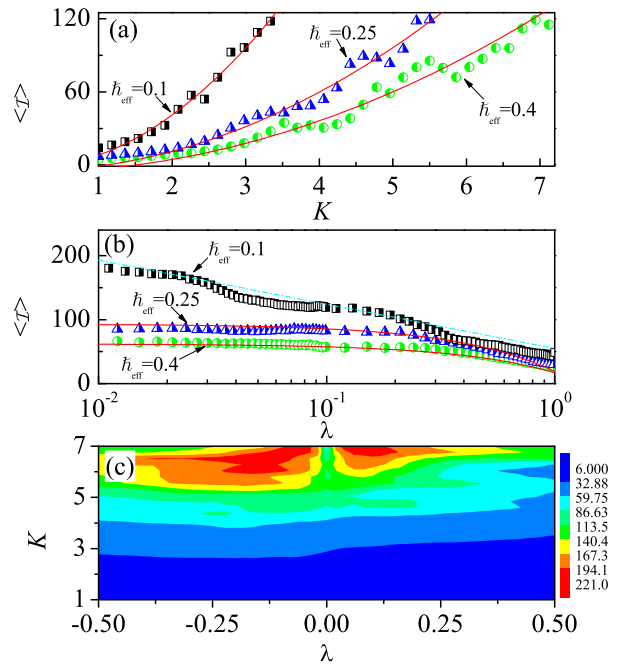}
\caption{Top two panels: $\langle\mathcal{I}\rangle$ versus $K$ (a) and $\lambda$ (b) with $\hbar=0.1$ (squares), 0.25 (triangles), and 0.4 (circles). In (a): Red lines indicate the function $\langle \mathcal{I}\rangle \propto \eta K^2$ with $\eta\approx 11$, 4.1, and 2.6 for $\ehbar=0.1$, 0.25, and 0.4, respectively. The parameter is $\lambda=0.003$. In (b): Dash-dotted line (in cyan) indicates the function $\langle\mathcal{I}\rangle\propto -\ln(\lambda)$. Solid lines in red denote $\langle\mathcal{I}\rangle\propto -\alpha \lambda$ with $\alpha\approx 74$ and 45 for $\ehbar=0.25$ and 0.4, respectively. The parameter is $K=5$. (c) The $\langle\mathcal{I}\rangle$ in the parameter space $(K,\lambda)$ with $\ehbar=0.25$.}\label{IPRQEStates}
\end{center}
\end{figure}

A commonly used quantity to measure the localization of a quantum state is the inverse participation ratio (IPR) $\mathcal{I}=(\sum_n|\psi_n|^2)^2/\sum_n|\psi_n|^4$\cite{Thouless72}. It is straightforward to prove that for an exponentially-localized state $|\psi_n|^2\sim e^{-n/\xi}$, the value of $\mathcal{I}$ is proportional to the localization length $\mathcal{I}\sim \xi$. In order to quantify the statistical feature of the localization of quasieigenstate, we numerically investigate the averaged value of IPR $\langle \mathcal{I}\rangle = \sum_{j=1}^{N}\mathcal{I}_j$, where $\mathcal{I}_j$ denotes the IPR of the $j$th quasieigenstate with $\varepsilon_i>0$. Figure~\ref{IPRQEStates}(a) shows that for a specific $\ehbar$, the $\langle \mathcal{I}\rangle$ increases in the quadratic function of $K$, i.e., $\langle \mathcal{I}\rangle\propto K^2$. In addition, the smaller $\ehbar$ is, the larger the $\langle \mathcal{I}\rangle$ is. This demonstrates the increase of localization length with the kicking strength or with the decrease of $\ehbar$, which is similar to the feature of the localization of quasieigenstates of Hermitian QKR~\cite{Izrailev90}.
We also numerically investigate the $\langle \mathcal{I}\rangle$ with varying $\lambda$. Figure~\ref{IPRQEStates}(b) shows that for small $\ehbar$ (e.g., $\ehbar=0.1$), the $\langle \mathcal{I}\rangle$ decays from a saturation level with increasing $\lambda$, in a logarithmic function $\langle \mathcal{I}\rangle\propto - \ln(\lambda)$. For larger $\ehbar$ (e.g., $\ehbar=0.25$ and 0.4), however, our numerical results support the linear decay $\langle \mathcal{I}\rangle\propto - \alpha \lambda$. Anyway, our investigation clearly reveals the enhancement of localization by increasing the non-Hermiticity, which is response for the decrease of the saturation value of $\langle p^2\rangle$ with the increase of $\lambda$ [see Figs.~\ref{Ener} and ~\ref{avEner}]. Figure~\ref{IPRQEStates}(c) shows the phase diagram of the $\langle \mathcal{I}\rangle$ for a wide regime of $K$ and $\lambda$. One can find the increase of $\langle \mathcal{I}\rangle$ with increasing $K$, and the decrease of $\langle \mathcal{I}\rangle$ with the increase of $\lambda$. In addition, the $\langle \mathcal{I}\rangle$ is asymmetric with the change $\lambda\rightarrow -\lambda$.

\section{Conclusion and discussion}\label{SEC-Conclusion}

In this work, we numerically investigate the dynamics of quantum irreversibility and energy diffusion in a NQKR model, for which the kicking strength is complex. We find that the exponential decay of Loschmidt echo $\mathcal{L}\propto e^{-2\gamma t}$ occurs only for very small $\lambda$. For sufficiently larger $\lambda$, the $\cal{L}$ remains at unity with time evolution, which demonstrates the disappearance of quantum irreversibility. The quantum diffusion exhibits the DL with time evolution, for which the saturation value of $\langle p^2\rangle$ decreases with increasing $\lambda$, signaling the enhancement of DL by non-Hermitian driven potential. The mechanism of quantum stability in the NQKR model is revealed by the fidelity between time-evolved quantum state and quasieigenstates. We find, both analytically and numerically, that a quantum state evolves to one of a quasieigenstate with significantly large $\varepsilon_i$, which is quantified by the emergence of maximum value of fidelity $\mathcal{F}(t_n)=|\langle\psi(t_n)|\varphi_{\varepsilon}\rangle|^2$ with a specific $\varepsilon_i$. Our numerical investigation on the averaged value of IPR shows the quadratic increase of $\langle \mathcal{I}\rangle$ with increasing $K$ and the decay of $\langle \mathcal{I}\rangle$ with increasing $\lambda$, which demonstrates the enhancement of exponential localization by non-Hermiticity.

Floquet-driven systems have now been accepted as ideal platforms for studying rich physics, such as many-body dynamical localization~\cite{Keser,Rozen,Tilen}, topological phase transition~\cite{Harper,LiangH,ZhouL}, and quantum thermalization~\cite{Fleckenstein}.
Understanding the quantum irreversibility and quantum diffusion of these systems has potential applications in the Floquet engineering on the propagation of optics in topological medium~\cite{JXu,Kumar,Quito} and the quantum transport in  twisted bilayer grapheme~\cite{Gabriel}. Quantum diffusion of matter receives intense attentions in different fields of physics ~\cite{Menu,Bitter,Paul}. Our finding of the destruction of exponential instability and the enhancement of DL by non-Hermitian driven potential shed light on the quantum diffusion in non-Hermitian chaotic systems.

\begin{acknowledgments}
This work was supported by the Natural Science Foundation of China (Grant No.12065009).
\end{acknowledgments}

\end{document}